\documentclass[%
 aip,
 amsmath,amssymb,
 reprint,%
]{revtex4-1}

\usepackage{graphicx}
\usepackage{dcolumn}
\usepackage{bm}
\usepackage{bbm}

\usepackage[utf8]{inputenc}
\usepackage[T1]{fontenc}
\usepackage{amsmath} 
\usepackage{amsthm}
\usepackage{appendix}
\usepackage{mathtools}
\usepackage{etoolbox}
\usepackage{comment}
\usepackage{lipsum}
\usepackage{svg}
\usepackage{color}
\usepackage{algorithm,algpseudocode}

\algnewcommand\algorithmicinput{\textbf{Input:}}
\algnewcommand\algorithmicoutput{\textbf{Output:}}
\algnewcommand\Input{\item[\algorithmicinput]}%
\algnewcommand\Output{\item[\algorithmicoutput]}%

\newcommand{\noop}[1]{}
\DeclareMathOperator{\Prob}{\mathbb{P}}

\newcommand{\br}{{\mathbf r}}

\makeatletter
\def\@email#1#2{%
 \endgroup
 \patchcmd{\titleblock@produce}
  {\frontmatter@RRAPformat}
  {\frontmatter@RRAPformat{\produce@RRAP{*#1\href{mailto:#2}{#2}}}\frontmatter@RRAPformat}
  {}{}
}%
\makeatother
\begin{document}

\preprint{AIP/123-QED}

\title[]{Reliable emulation of 
complex functionals by active learning with error control}
\author{Xinyi Fang}
\affiliation{Department of Statistics and Applied Probability, University of California, Santa Barbara, CA 93106, USA}
\author{Mengyang Gu}%
\homepage{Corresponding author: \href{mailto:mengyang@pstat.ucsb.edu}{mengyang@pstat.ucsb.edu}
}
\affiliation{Department of Statistics and Applied Probability, University of California, Santa Barbara, CA 93106, USA}
\author{Jianzhong Wu }
\homepage{Corresponding author: \href{mailto:jwu@engr.ucr.edu}{jwu@engr.ucr.edu}
}
\affiliation{ 
Department of Chemical and Environmental Engineering, University of California, Riverside, CA 92521, USA
}%

\date{\today}

\begin{abstract}
A statistical emulator can be used as a surrogate of complex physics-based calculations to drastically reduce the computational cost. 
Its successful implementation hinges on an accurate representation of the nonlinear response surface with a high-dimensional input space. Conventional ``space-filling'' designs, including random sampling and Latin hypercube sampling, become inefficient as the dimensionality of the input variables increases, and the predictive accuracy of the emulator can degrade substantially for a test input  away from  the training input set. To address this fundamental challenge,  we develop a reliable emulator for predicting complex functionals by active learning  with error control (ALEC).  
The algorithm is applicable to infinite-dimensional mapping with high-fidelity predictions and a  controlled predictive error. The computational efficiency has been demonstrated by emulating the classical density functional theory (cDFT) calculations, a statistical-mechanical method widely used in modeling the equilibrium properties of complex molecular systems. 
We show that ALEC is much more accurate than conventional emulators based on the Gaussian processes  with ``space-filling'' designs and alternative active learning methods. Besides, it is computationally more efficient than direct cDFT calculations. ALEC can be a reliable building block for emulating expensive functionals owing to its minimal computational cost, controllable predictive error, and fully automatic features.  
\end{abstract}
\maketitle

\section{Introduction}
Theoretical research in chemistry and materials science is often hampered by computational bottlenecks in applying complex physical models to predict the microscopic structure and physicochemical properties of many-body systems. Statistical and machine learning (ML) methods, such as Gaussian process (GP) regression, basis expansion methods, and neural network models, have been widely used as a surrogate of complex physical models to emulate the atomic force fields, density functionals, chemical reactions, and diverse properties of materials and chemical systems \cite{bartok2010gaussian,chmiela2018towards,shapeev2016moment,schutt2018schnet,lin2020analytical,li2021kohn,deringer2021gaussian,yatsyshin2022physics}. 
Here a typical aim of a ML approach is to learn the map with a high dimensional input or a functional input from observations. 
Trained on a set of pre-specified inputs, a statistical emulator can obtain accurate predictions when a test point is close to the training inputs
\cite{westermayr2021perspective, unke2021machine,sajjan2022quantum}.

Constructing a statistical emulator often starts with selecting a set of ``space-filling'' samples in the input space, such as  uniform samples or Latin hypercube samples (LHS) \cite{santner2003design}. The statistical emulator will be trained based on the outputs  of the simulation at these pre-selected inputs. The idea is to ensure that for each test point in the input space, there are a sufficiently large number of inputs near this test point, where the outcomes (such as the potential energy or atomic force in emulating the first-principles calculations) were observed. Using the space-filling samples and assuming a set of regularity conditions, the predictive error of some ML methods, such as the GP regression, is guaranteed to converge to zero,  
when the sample size goes to infinity \cite{van2009adaptive}. 
The space-filling samples have two main drawbacks in emulating physics-based calculations such as molecular dynamics simulation (MD). First, the input values, such as atomic positions or pairwise inter-atomic distances in MD, may not be directly controlled.
Second, physics-based modeling of molecules and materials often involves a high or infinite dimensional input space, such as the  external potential function and atomic configurations. If the statistical emulator is trained  in a small parameter subspace, the  predictive accuracy of the emulator can be dramatically degraded outside the subspace where no training input is available. 
This scenario often occurs when the outcome is required in a slightly different system or at another  experimental condition, such as different temperatures or pressures, whereas the statistical emulator becomes inaccurate if it is only trained on one set of conditions. Obtaining accurate predictions of any test input with a controlled predictive error is important to bypass expensive computations based on complex physical models.

To address the fundamental difficulty in predicting functions and functionals with a high-dimensional input space, 
active learning has become one of 
the most promising techniques to sequentially reinforce the emulation of physics-based modeling \cite{schutt2020machine}. 
The active learning approach can be implemented in two broad scenarios. The first one is an online scenario or ``on-the-fly'' prediction, where the test inputs sequentially come and one does not know what future input needs to be predicted \cite{podryabinkin2017active, uteva2018active, vandermause2020fly}. The online prediction scenario has many applications, such as  Bayesian optimization \cite{shahriari2015taking,shields2021bayesian} and model calibration  \cite{kennedy2001bayesian,gu2018sgasp},  where the outcome of the simulation  needs to be sequentially predicted for every sampled input. 
The second scenario is to sequentially select the samples to run a computer simulation for predicting the entire input space, where the uncertainty of the emulator is often used for selecting the next design run \cite{sauer2022active}. In this work, we focus on the online prediction scenario and test our approach using the classical density functional theory (cDFT) calculations for one-dimensional (1D) hard-rod systems, with different external potentials. The availability of exact densities and free-energy functional for various 1D systems of hard rods allows us to validate the efficiency of the surrogate model with the ground truth. Here the fundamental difficulty is on a user-specified external potential--an arbitrary functional input with infinite dimensions. For constructing statistical emulators, we assume that no parametric form of this functional input is available.  We demonstrate that the new integrated approach provides a high-fidelity prediction of particle density profiles and grand potentials with a controlled error in the probabilistic sense, more accurate than a GP emulator with conventional ``space-filling'' designs, and it is more computationally scalable than direct cDFT calculations.


We highlight the key novelty of this work in developing the surrogate model and error control  criteria for functional mapping. First, while some popular criteria, such as the D-optimality, were introduced in the active learning framework \cite{podryabinkin2017active,yue2020active}, the threshold of determining whether a test input can be reliably predicted may be hard to choose, as the threshold may not be directly related to the  error in prediction. In practice, one may need to tune the threshold for different systems   for a given error bound.   Here, we use the internal uncertainty assessment of the GP emulator to decide whether a test input can be predicted precisely. 
For any given threshold of predictive error and probability tolerance bound, our approach automatically defines the criterion to control the predictive error below the threshold with a large probability satisfying the probability tolerance bound (see Section \ref{subsec:error_control}). 
We demonstrate that our  approach can identify whether an upcoming input can be reliably predicted,  and thus it has much better performance than the conventional statistical emulation by training the emulator with a space-filling design, such as random sampling. Furthermore, we derive an iterative formula that reduces the computational complexity for Gaussian process models with augmented samples. In each iteration, updating our algorithm only requires $\mathcal O(n^2)$ operations, while a direct approach requires  $\mathcal O(n^3)$ operations. 
Other emulation approaches, such as those enforcing the invariance symmetries, can be easily included in our integrated framework for reliable predictions with a controlled error. 

The remainder of the paper is structured as follows: In Section \ref{sec:cdft}, we provide a brief overview of the classical density functional theory in the context of the 1-dimensional (1D) hard-rod model used in this work. In Section \ref{sec:ppgp}, we introduce the parallel partial Gaussian process for emulating physics-based calculations with vectorized output on massive grid points, such as particle densities from cDFT calculations.
Then, in Section \ref{subsec:error_control}, we introduce our proposed method to control the predictive error. The computationally scalable update of the algorithm and its computational cost are discussed in Sections \ref{subsec:update_derivation} and \ref{subsec:algorithm_cost}, respectively. 
The numerical comparison in Section \ref{sec:num_results} demonstrates that our approach outperforms the conventional GP emulators with random-sample designs and the active learning approach with the D-optimality criterion. We conclude the paper along with perspectives on some future research directions in Section \ref{sec:conclusion}.

\section{Classical density functional theory}
\label{sec:cdft}
\begin{figure*}[ht]
    \centering
    \includegraphics[width=16.5cm]{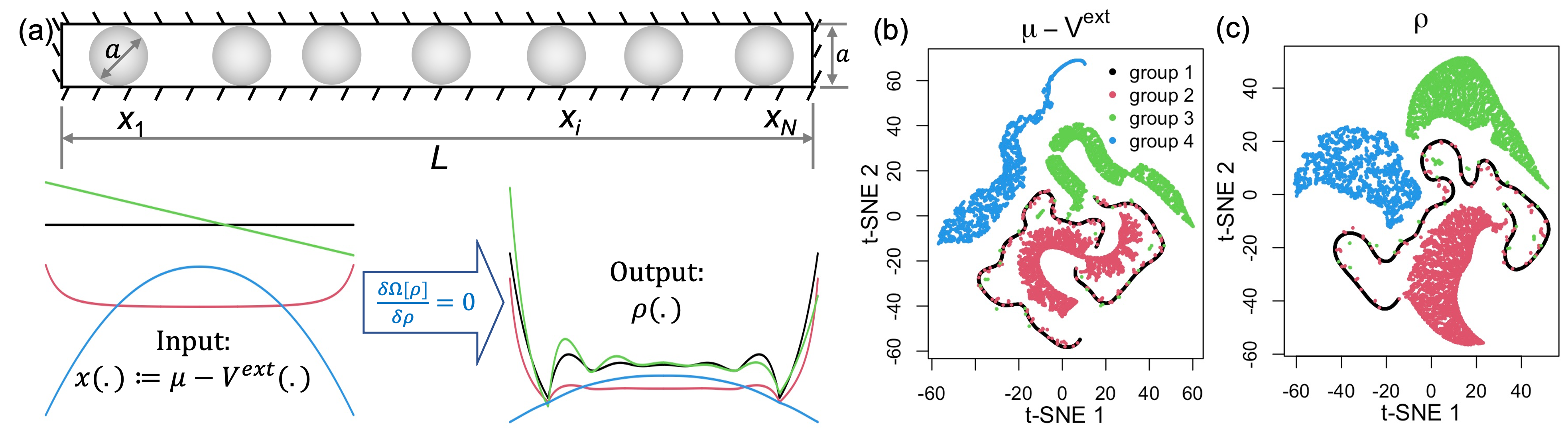}
    \caption{
    (a) Functional learning in the context of cDFT for one-dimensional hard rods of uniform size $a$ in an external field $V^{ext}(.)$ defined in a domain of size $L$. (b)-(c) t-distributed stochastic neighbor embedding (t-SNE) for visualization of input $\mu- V^{ext}(.)$ (middle panel) and output $\rho(.)$ (right panel). For each group,  $2000$ density profiles are generated based on different combinations of $\mu$ and parameters in $V^{ext}(.)$.}
    \label{fig:functional_learn}
\end{figure*}
Classical density functional theory (cDFT) is a statistical-mechanical method that describes the thermodynamic properties of equilibrium systems in terms of the density profiles of individual particles \cite{Evans92, Wu07}. In a nutshell, cDFT calculations are based on the minimization of   
a grand potential functional that, for systems consisting of only one type of particle, can be expressed as
\begin{align}
    \Omega[\rho] =   F_{id}[\rho]+ F_{ex}[\rho]+\int d \br \rho(\br)[V^{ext}(\br)-\mu],
    \label{equ:Omega}
\end{align}
where $\rho(\br)$ denotes the particle density at position $\br$, $F_{id}[\rho]$ is the free-energy functional of an ideal gas, and $F_{ex}[\rho]$ is called the excess free-energy functional, $V^{ext}(\br)$ denotes a one-body external potential, and $\mu$ is the chemical potential of the particles. The ideal-gas functional has an exact form 
\begin{equation}
     \beta F_{id}[\rho] = \int d\br \rho(\br)\{\ln[\rho(\br)\Lambda^3]-1\},
\end{equation}
where $\Lambda$ is the thermal wavelength and $\beta=1/(k_BT)$ with $k_B$ being the Boltzmann constant and $T$ the absolute temperature. While $T$ and $\mu$ are thermodynamic variables that define the equilibrium condition, $\Lambda$ and $F_{ex}[\rho]$ depend on the intrinsic properties of the system such as the particle mass and inter-particle potential. In this paper, we set $\beta=\Lambda=1$ as the units of energy and length.

One key premise of cDFT calculations is that $V^{ext}(\br)$ is uniquely determined by $\rho(\br)$ such that, given an analytical expression for $F_{ex}[\rho]$, $\rho(\br)$ can be solved by minimizing $\Omega[\rho]$ and,  subsequently, all thermodynamic properties of the system can be derived. The equilibrium density profile minimizes $\Omega$ and thus satisfies the Euler-Lagrange equation: 
\begin{align}
    \rho(\br) = \exp \big\{ \mu - {\delta F_{ex}}/{\delta\rho}- V^{ext}(\br) \big\}.
    \label{equ:rho_EL}
\end{align}
Eq.\eqref{equ:rho_EL} provides a functional map between the one-body density $\rho(\br)$ and the one-body external potential $V^{ext}(\br)$. Because in general $F_{ex}$ is a complicated functional of $\rho(\br)$, solving $\rho(\br)$ from the Euler-Lagrange equation is often computationally demanding. Besides, the formulation of an accurate expression for $F_{ex}$ is a formidable task for complex molecular systems.  In this work, we demonstrate that statistical emulation will be helpful to solve both problems. 

For proof of concept, we consider a statistical emulator of cDFT for one-dimensional (1D) systems consisting of identical hard rods (HR). For such systems, the excess free-energy functional, thus the grand potential, is exactly known \cite{percus1976equilibrium}. The analytical results were derived decades ago and are reproduced in Appendix \ref{sec:close_form_cdft}. Given any external potential $V^{ext}(\br)$ and chemical potential $\mu$, the density profile of hard rods can be numerically solved (e.g., by Picard iteration) from Eq. (\ref{equ:rho_EL}), and it is plugged into Eqs. (\ref{equ:rho_EL}),  (\ref{equ:close_Omega}), and (\ref{equ:close_F_ex}) for computing the grand potential and other thermodynamic quantities. 


For a given system defined by $\mu$ and $ V^{ext}(\br)$, statistical emulation aims to predict the particle density and free-energy functionals as in direct cDFT calculations (illustrated in Fig. \ref{fig:functional_learn}(a)). 
Previous work in cDFT \cite{shang2019classical,lin2020analytical,cats2021machine} and Kohn-Sham density functional theory (KS-DFT) \cite{snyder2012finding, brockherde2017bypassing} demonstrate the performance of ML models with a single parametric form for the external potential with different choices of parameters. In this work, we show that the active learning procedure is applicable to multiple classes of external potential similar to cDFT calculations with the same intrinsic Helmholtz energy functional. We demonstrate that, when presented with a new class of external potentials, the model can discern whether this new input can be accurately predicted. 
 As mentioned above,  functional mapping represents one of the biggest obstacles in statistical emulation as well as machine learning. 
Filling the entire functional input space directly with a limited number of simulation runs, such as random sampling or the LHS, is not generally possible. When the external potential function is distant from the training input set, both statistical emulators and machine learning methods become inaccurate. 
In this work, we provide a solution to this fundamental problem by integrating direct cDFT calculations 
with a Gaussian process (GP) emulator through an active-learning framework. 
We derive a probabilistic bound to control the overall prediction error of the particle density
based on the internal assessment of the uncertainty from the statistical emulator, making our approach 
applicable to functional learning in infinite dimensional spaces, as well as reducing the computational costs for inputs that can be accurately predicted. Here our algorithm can reliably learn multiple classes of external potentials, while earlier applications of ML methods in both KS-DFT (e.g. \cite{brockherde2017bypassing}) and cDFT (e.g. \cite{lin2020analytical,cats2021machine}) only demonstrate the performance of their models on one particular class of the external potential. Besides, the active learning scheme developed in this work enables the ML approaches to discern whether outcomes of a test input can be reliably predicted. This ability allows us to control predictive errors as well as substantially reduce the computational cost.

In Fig. \ref{fig:functional_learn}(b) and (c), the input $\mu- V^{ext}(.)$ and output $\rho(.)$ of the statistical emulator are projected onto two t-distributed Stochastic Neighborhood Embedding (t-SNE) factors as commonly used for visualizing high-dimensional data \cite{van2008visualizing}. It is evident that the inputs generated from different groups of input functions form different clusters. Similar patterns can be observed for the corresponding particle densities. 
In addition, group 1 forms a curve in Fig. \ref{fig:functional_learn}(b) due to the fact that the only change of the input in group 1 is $\mu$,  which only results in a shift in the input space, making group 1 mapping the simplest to learn for the statistical emulator, as to be demonstrated later in the section on numerical results. 
We noticed several points in groups 2 and 3 are close to those in group 1. These points correspond to  samples whose input parameters are close to 0 and are thus nearly equivalent to group 1's inputs. These patterns demonstrate the input-output relationship can be captured by distance metrics, as the smaller distance between chemical potentials leads to the higher similarity between the particle densities.  This validates the use of kernels in the Gaussian process to map the distance between inputs (chemical and external potentials) to the correlation between output (particle densities), as the smaller distance in the input space indicates a larger correlation or similarity between outputs. Unlike the dimension reduction approach shown in Fig. \ref{fig:functional_learn}, we will model the entire density profile jointly in a statistical emulator to encode all information. 

\begin{figure}[t]
    \centering
    \includegraphics[width=7.5cm]{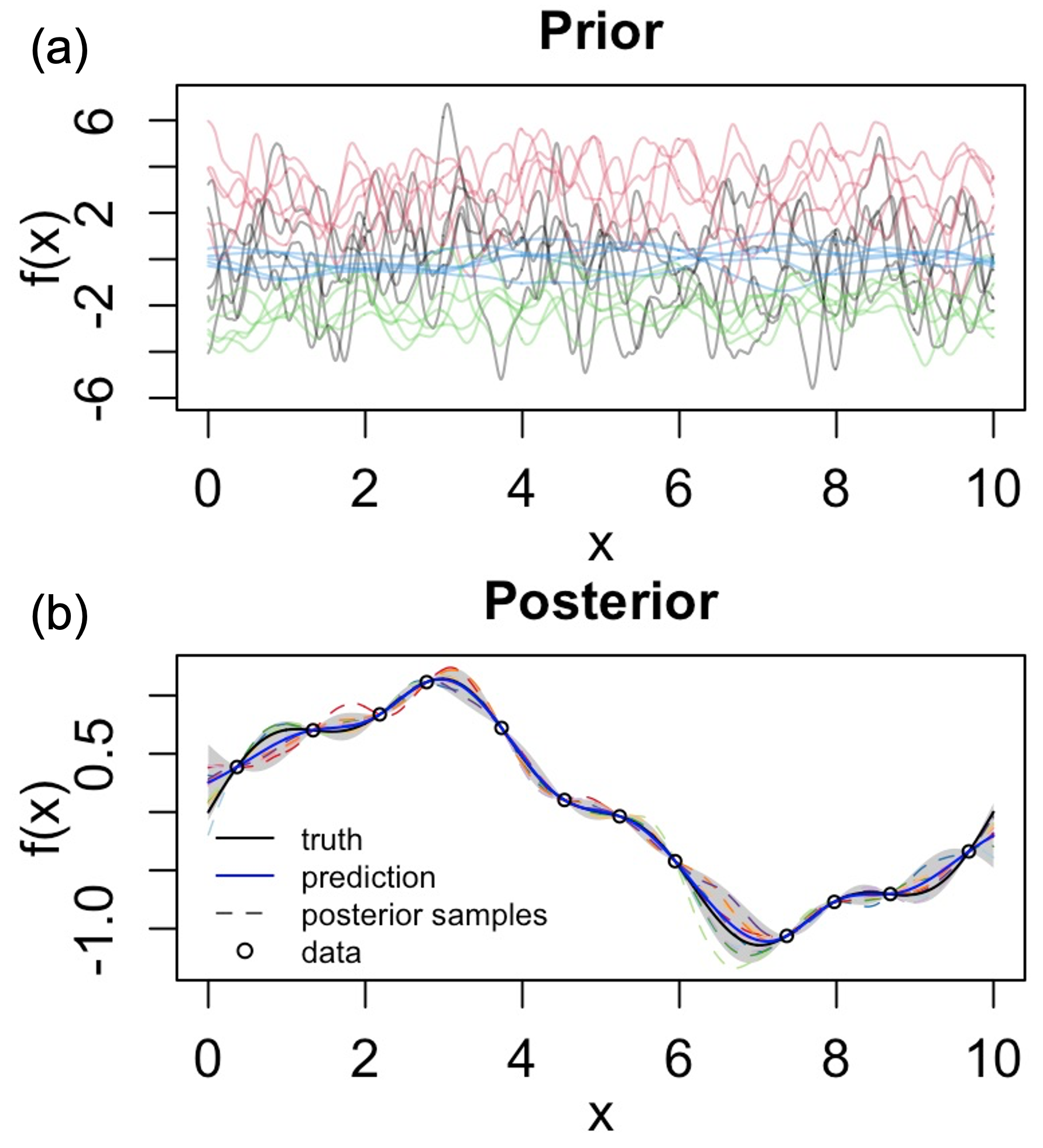}
    \caption{Illustration of Bayesian updating in a GP emulator. (a) Prior samples from 4 groups (indicated by different colors) with constant mean and Mat{\'e}n kernel in \eqref{equ:matern_5_2}. The different parameters are specified as  $\gamma = 0.2, \sigma = 2, \mu = 0$ (black curves); $\gamma = 0.25, \sigma = 1.5, \mu = 3$ (red curves);  $\gamma = 0.33, \sigma = 1, \mu = -2$ (green curves);  $\gamma = 1, \sigma = 0.5, \mu = 0 $ (blue curves). (b)  The truth (black curve), the predictive mean (blue curve), the 95\% predictive credible interval (grey region), and 10 posterior predictive samples (colored dashed lines) after observing 12 data points (black circles).}
    \label{fig:gp_prior_post}
\end{figure}
\section{A reliable Gaussian process emulator by active learning with error control}
\label{sec:active_learning}
\subsection{Parallel partial Gaussian process emulation for vectorized outputs}
\label{sec:ppgp}

For demonstration purposes, our statistical emulator is built upon a parallel partial Gaussian Process (PP-GP) emulation approach  \cite{Gu2016PPGaSP} for  representing the map between the particle density, a vectorized output on massive grid points, and the input function from the chemical and external potentials. Other emulators that produce  uncertainty quantification for predictions can be used as well.  
The GP emulator is one of the most commonly used surrogate models for expensive physics-based calculations, and it has been widely used for emulating molecular simulation, the potential energy surface, and atomic force fields
\cite{bartok2010gaussian,rupp2012fast,chmiela2017machine, li2022efficient,deringer2021gaussian}. In combination with the dimension reduction technique, the GP  emulator has also been used as a surrogate model for approximating the KS-DFT calculation of electron density \cite{brockherde2017bypassing}.

Fig. \ref{fig:gp_prior_post} illustrates the Bayesian updating formulation in GP from prior to posterior with one-dimensional inputs and outputs. 
As shown in Fig. \ref{fig:gp_prior_post}(a), the GP prior is a very flexible class of functions, that include smooth or wiggly functions with large or small fluctuation, using different choices of parameters. After assuming 12 Latin hypercube samples from a function $f(x) = \sin(2\pi x/10)+0.2\sin(2\pi x/2.5)$, in Fig. \ref{fig:gp_prior_post}(b), we plot  the predictive mean from GP (blue curve) and the actual value of $f(x)$ (black curve), as well as 10 predictive posterior samples that fit the data points with optimized parameters and the 95\% predictive interval (gray area), using a GP emulator implemented in \cite{gu2018robustgasp}, where the parameters are estimated by the posterior mode. First, we found a small number of samples can constrain the predictive mean and posterior samples to be around the truth. Second, the truth are almost all covered by the 95\% predictive interval, indicating appropriate quantification of the predictive uncertainty. 
This internal uncertainty assessment by the GP emulator enables us  to develop the criterion to control the predictive error for emulating functions with high-dimensional inputs in a probabilistic way, which will be illustrated in section \ref{subsec:error_control}. 

We use the PP-GP emulator to model the particle density in cDFT. 
To highlight the salient feature of the PP-GP emulator, 
let $\mathbf x = \mu- V^{ext}(.)$ denote the functional input discretized at $p$ spatial grid points. For any input $\mathbf x$, the particle density is defined at $k$ grid points, written as $\bm\rho(\mathbf x)=[\rho_1(\mathbf x), \rho_2(\mathbf x), ..., \rho_k(\mathbf x)]$. The number of spatial grid points used in discretizing external potential and density can be different but, for simplicity, here we let $p=k$.  Given any $n$ cDFT calculations with inputs $\{\mathbf x_1,...,\mathbf x_n\}$, the density vector at any spatial grid point is assumed to follow a multivariate normal distribution  $\bm\rho_j = [\rho_j(\mathbf x_1), \rho_j(\mathbf x_2), ..., \rho_j(\mathbf x_n)]^T \sim \mathcal {MN}(\bm \mu_j, \sigma_j^2\mathbf R),$ where
$\bm \mu_j=[\mu_j(\mathbf x_1),...,\mu_j(\mathbf x_n)]^T$ is the mean vector, $\sigma^2_j$ is a variance parameter, and $\mathbf R$ is an $n\times n$ correlation matrix between density at any grid point over $n$ input scenarios, with $(i,j)$ entry parameterized by a kernel function $K(\mathbf x_i,\mathbf x_j)$.

At any input $\mathbf x$, the mean of the output is commonly modelled as  $\mu_j(\mathbf x) = \mathbf h(\mathbf x)\bm\theta_j$, where $\bm \theta_j$ is a $q$-dimensional vector, and $\mathbf h(\mathbf x) = [h_1(\mathbf x), ..., h_q(\mathbf x)]$ is a row vector of mean basis function. Here we only use the intercept, i.e. $h(\mathbf x)=1$ and $q=1$. Note that the mean parameter and variance parameter are distinct for each spatial grid point of the particle density. Having different mean and variance parameters makes the model more flexible to capture the variability of particle densities at different grid points.

We assume that the kernel function is isotropic, i.e., it is a function of the Euclidean distance between any pair of inputs. The correlation  is commonly modeled by  
the power exponential kernel functions or the Mat{\'e}rn kernel functions \cite{rasmussen2006gaussian}. 
In this work, we use the Mat{\'e}rn kernel function with roughness parameter $5/2$ to model the correlation between input $\mathbf x_a$ and $\mathbf x_b$: 
\begin{equation}
    K(\mathbf x_a,\mathbf x_b) = \left(1+\sqrt{5}\frac{d}{\gamma}+\frac{5d^2}{3\gamma^2}\right)\exp\left(-\sqrt{5}\frac{d}{\gamma}\right),
    \label{equ:matern_5_2}
\end{equation}
where $d=||\mathbf x_a-\mathbf x_b||$, with $||\cdot||$ denoting the Euclidean distance, and $\gamma$ is a range parameter that can be estimated by the training data. The sample path of the GP with the Mat{\'e}rn kernel function in (\ref{equ:matern_5_2}) is twice mean differentiable  \cite{rasmussen2006gaussian}, assuring good predictive performance in emulating smooth functions. 

Assume we have run cDFT calculations at $n$ different input scenarios $\{\mathbf x_1,...,\mathbf x_n\}$, leading to a $k\times n$ particle density matrix $\bm \rho=[\bm \rho_1,..., \bm \rho_n]$. 
We compute the predictive distribution after integrating out the mean and variance parameters for making predictions with the reference prior of the mean and variance parameters  \cite{berger2001objective},  $\pi(\bm \theta_j,\sigma^2_j)\propto 1/\sigma^2_j$, for spatial grid point $j=1,...,k$.  Conditional on estimated range parameter $\hat\gamma$ and output density at the $j$th coordinate $\bm\rho_j$, the predictive distribution of $\rho_j(\mathbf x^*)$ at any new $\mathbf x^*$ and coordinate $j$, follows a non-centered scaled Student's t-distribution with $n-q$ degrees of freedom:
\begin{equation}
    \rho_j({\mathbf x^{*}})| \hat\gamma, \bm \rho_j \sim \mathcal {T} ( \hat \rho_j({\mathbf x}^{*}),\hat{\sigma}_j^2K^{**}, n-q),
\label{equ:predictiongp}
\end{equation}
where 
\begin{align}
    \hat{\rho}_j ({\mathbf x}^{*}) &= { \mathbf h({\mathbf x}^{*})} \hat{\bm{\theta}}_j+\mathbf{r}^T(\mathbf{x}^*){{\mathbf R}}^{-1}\left(\bm{\rho}_j-\mathbf H\hat{\bm{\theta}}_j\right), \label{equ:gppredmean}\\
    K^{**} &= K^{*} +  {\mathbf  h}^*(\mathbf x^*)^T \left(\mathbf H^T{{\mathbf R}}^{-1}\mathbf H \right)^{-1} {\mathbf  h}^*(\mathbf x^*) ,  \label{equ:gppredcorrelation}
\end{align}
with  $ {\mathbf h}^*(\mathbf x^*)={{\mathbf h(\mathbf x^{*})}}-\mathbf H^T {{\mathbf R}}^{-1}\mathbf{r}(\mathbf{x}^*)$, and 
\begin{align}
    \hat{\bm{\theta}}_j&=\left( \mathbf H^T {{\mathbf R}}^{-1} \mathbf H \right)^{-1}\mathbf H^T{{\mathbf R}}^{-1}\bm\rho_j,\\
    \hat{\sigma}_j^2 &= {(\bm\rho_j-\mathbf H \hat{\bm{\theta}}_j)}^{T}{{\mathbf R}}^{-1}({\bm\rho_j}-\mathbf H\hat{\bm{\theta}}_j)/(n-q). \label{equ:sigma_j_hat}
\end{align}
Here the $n\times q$ input mean basis matrix follows $\mathbf H=[\mathbf h^T(\mathbf x_1),...,\mathbf h^T(\mathbf x_n)]^T$, and $K^*=K({\mathbf x^{*}}, {\mathbf x^{*}})-{ \mathbf{r}^T(\mathbf{x}^*){ {\mathbf R}}^{-1}\mathbf{r}(\mathbf{x}^*)}$ denotes the correlation between test input and training input, with  $\mathbf{r}(\mathbf{x}^*) = [K(\mathbf{x}^*,{\mathbf{x}}_1 ), \ldots, K(\mathbf{x}^*,{\mathbf{x}}_n )]^T$.   
The derivation of Eq.  (\ref{equ:predictiongp}) can be found in the supplement materials. In PP-GP, the distribution of the output is assumed to be independent at two grid points, and thus we have $p(\rho_j({\mathbf x^{*}})\mid \hat\gamma, \bm \rho)=p(\rho_j({\mathbf x^{*}})\mid \hat\gamma, \bm \rho_j)$, for any $j$ and $\mathbf x^*$, which allows us to use Eq.  \eqref{equ:predictiongp} for computing the posterior predictive distribution. For further justification of the assumption, see Theorem 6.1 in \cite{Gu2016PPGaSP}.   
The  PP-GP emulator was implemented in the "RobustGaSP" package \cite{gu2018robustgasp}, where the marginal posterior mode estimator is used as a default method to estimate the range parameter $\gamma$ \cite{Gu2016PPGaSP}.  

In comparison with alternative methods, the PP-GP emulator discussed above has advantages in terms of computational scalability and internal uncertainty assessment. First, constructing the covariance matrix requires $O(n^2p)$  operations, while computing the predictive mean in Eq. (\ref{equ:gppredmean}) only takes $\mathcal O(nk)+\mathcal O(n^3)$ operations, where $n$, $p$, and $k$ denote the number of training simulation runs, the number of discretization points in the input functions and particle densities, respectively.
The $\mathcal O(n^3)$ operations come from the Cholesky decomposition for computing the  inversion and determinant of the covariance matrix.   In contrast, building a separate emulator independently of particle densities at each grid point takes $\mathcal O(n^3k)$ operations for computing the predictive mean, due to the inversion of the distinct covariance matrices at each spatial grid point. Approaches that model the spatial covariance matrix generally take $\mathcal O(k^3)$ operations, which is even slower as the number of spatial grid points $k$ of the density is large. Second, the uncertainty in predictions can be directly assessed as  any quantiles and percentiles of the predictive distributions in (\ref{equ:predictiongp}) have closed-form expressions. The assessed uncertainty allows us to  control the predictive error through integrating a numerical solver and an emulator into the algorithm, discussed below.   

\subsection{ALEC: active learning with error control approach}
\label{subsec:error_control}

Active learning is a sequential design approach in statistics. In an ``on-the-fly'' online prediction scenario, there is no information regarding the future input to be predicted. If a configuration is new to the current training set, i.e., it is located in a sparsely sampled region, we may not be able to accurately predict this sample by the emulator. For a large  functional input space, it is almost inevitable to extrapolate the sampled input space, as the number of simulated runs that populate this input space is sparse.  Therefore, it is necessary to establish a criterion for determining whether an upcoming configuration can be reliably predicted. 
\begin{figure}[t]
    \centering
    \includegraphics[width=8.6cm]{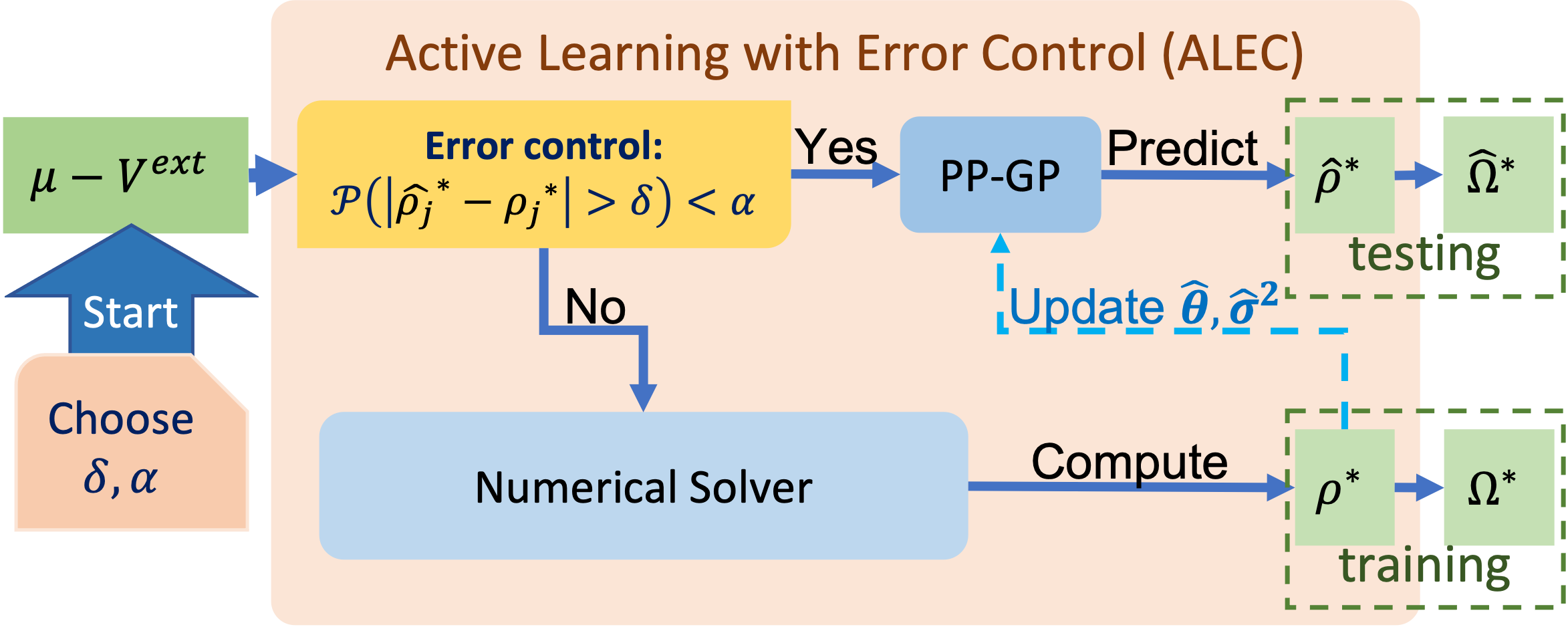}
    \caption{
    An overview of the active learning with error control framework. 
    }
    \label{fig:workflow}
\end{figure}

In this work, we develop the active learning with error control (ALEC) emulator that can automatically control predictive error for simulations with high or infinite dimensional input space. 
A few criteria  were studied before to detect the test case hard to be predicted, such as the D-optimality criterion \cite{podryabinkin2017active} and the predictive variance criterion \cite{vandermause2020fly}. Yet, the threshold of these strategies often requires to be chosen empirically, as they may not be directly related to the predictive error. In our automatic approach, a new criterion has been established such that the threshold can be chosen automatically for any given predictive error bound and probability tolerance bound, based on the internal uncertainty assessment of the emulator. Fig. \ref{fig:workflow} shows an overview of our strategy. For a given testing input $\mathbf x^*$, if 
the criterion that controls the predictive error is satisfied as shown in the yellow box, the particle density profile will be predicted; otherwise, a numerical solver will be called for computing the particle density corresponding to $\mathbf x^*$, and this new information will be added to the training set and used to update the PP-GP emulator.  


\begin{figure*}[t]
    \centering
    \includegraphics[width=16.5cm]{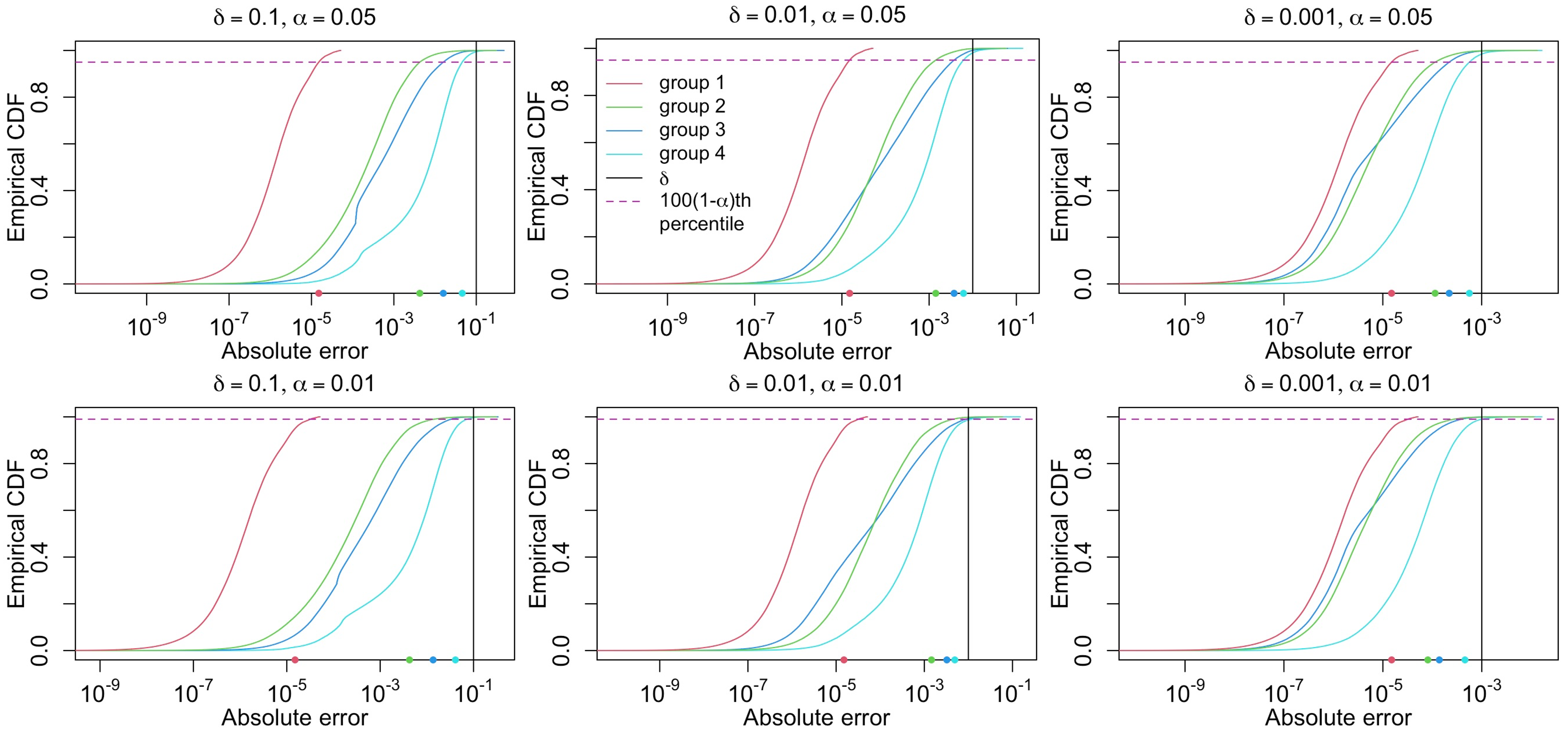}
    \caption{Empirical CDF of absolute error of particle densities $|\rho_j(\mathbf x^*_i)-\hat\rho_j(\mathbf x^*_i)|$  for 4 classes of input functions at all grid points using the criterion (\ref{equ:criterion}) with different error bounds $\delta$ and probability tolerance thresholds $\alpha$. The  horizontal line and the vertical line are $1-\alpha$ and $\delta$, respectively.  On the x-axis,  the $1-\alpha$ percentiles of the absolute errors  are recorded, all of which are less than $\delta$. 
    }
    \label{fig:error}
\end{figure*}

Let us first consider controlling the predictive error for particle density at a fixed coordinate $j$ based on the PP-GP algorithm. Given a prespecified error bound $\delta_j>0$,  we aim to have the predictive error of a test input $\mathbf x^*$ smaller than $\delta_j>0$ for more than $1-\alpha$ of the predictions, where $\alpha$ is a small value (e.g. $\alpha=0.05$).  This means  
\begin{align}
\mathbb P \big(|\rho_j(\mathbf x^*)-\hat{\rho}_j(\mathbf x^*)|>\delta_j\big) \leq \alpha,
\label{equ:P_error_j}
\end{align}
i.e., the probability that the absolute error between the predicted and true densities at the $j$th coordinate exceeds $\delta_j$ is less than $\alpha$. 
Since small predictive variance from the emulator indicates high-fidelity in prediction, we predict  the test input $\mathbf x^*$ if
\begin{equation}
    \sqrt{\hat\sigma^2_j K^{**}}\leq \frac{\delta_{j} }{t_{\alpha/2}(n-q)},
    \label{equ:jth_criterion}
\end{equation}
where $t_{\alpha/2}(n-q)$ is the upper $\alpha/2$ quantile of a Student's t-distribution with $n-q$ degrees of freedom,  $K^{**}$ and $\hat \sigma^2_j$  are given in Eqs. (\ref{equ:gppredcorrelation}) and (\ref{equ:sigma_j_hat}), respectively. 
Otherwise, we will compute this particle density directly from cDFT using a numerical solver, as the assessed uncertainty for predicting this test input is large. This strategy satisfies the probabilistic error control outlined in Eq. (\ref{equ:P_error_j}), as shown in 
Appendix \ref{sec:prob_ub}.

We have a few remarks regarding this strategy. 
First, the error bound $\delta_j$ and probability tolerance $\alpha$ have interpretations. Suppose, for instance, one expects to see the predictive error in more than $95\%$ of predictions smaller than $0.01$, one can then let $\delta=0.01$ and $\alpha=0.05$. 
Second, in the active learning strategy, one reduces the sample space from the original space $\mathcal X$ to $ \mathcal X_{F,j}=\{\mathbf x^*: \sqrt{\hat\sigma^2_j K^{**}}\leq \delta_{j}/t_{\alpha/2}(n-q) \}$, where the subscript $F$ denotes the input set that can be reliably predicted. A numerical solver for cDFT needs to be called for any input $\mathbf x^* $ that does not satisfy Eq. (\ref{equ:jth_criterion}).  Though selecting a smaller  $\delta$ or $\alpha$  leads to a smaller predictive error, more evaluations from the numerical solver would be required, which increases the computational cost. Thus, a balance between the size of the cut-off value and computational time is needed in practice. 
Third, the degree of freedom of the Student's t-distribution increases with the sample size. When the sample size is sufficiently large, the t-distribution can be accurately approximated by a standard normal distribution.  For instance, when $n-q=300$ and $\alpha=0.05$, $t_{\alpha/2}(n-q)\approx 1.968$ and the upper $\alpha/2$ quantile of a normal distribution $Z_{\alpha/2}\approx 1.960$, which only differs from that of a t-distribution by around $0.5\%$. Thus one can use the normal approximation if the sample size is large. Finally,  the error control from Eq. (\ref{equ:jth_criterion}) 
means that if the statistical emulator is a correct representation of the reality, and the number of test samples is  infinitely large, we have  at least $1-\alpha$ probability such that the absolute predictive error is less than $\delta$.  This does not preclude the cases where in slightly more than $1-\alpha$ of the predictions, the predictive error is larger than $\delta$, due to model misspecification or the limited number of test samples.  Fortunately,   the predictive error at these inputs is typically close to $\delta$, as the assessed uncertainty of these inputs is smaller than the threshold.

We can extend the strategy for controlling the error of the density at a fixed coordinate $j$ to the overall error of density prediction. Consider that we use a uniform cutoff value $\delta$, and that we make predictions for a test input $\mathbf x^*$ if each coordinate satisfies Eq. (\ref{equ:jth_criterion}), i.e.,
\begin{equation}
    \sqrt{\max_j\{\hat\sigma^2_j K^{**}\}}\leq \frac{\delta}{t_{\alpha/2}(n-q)}.
    \label{equ:criterion_max}
\end{equation}
We call the criterion in Eq. \eqref{equ:criterion_max} the maximum variance selection criterion. Under this criterion,  we are able to derive the probabilistic bound to control the predictive error of particle density at each coordinate, 
\begin{equation}
    \mathbb P \big(|\rho_j(\mathbf x^*)-\hat{\rho}_j(\mathbf x^*)|>\delta\big) \leq \alpha,  \quad \text{for } j=1,...,k,     \label{equ:P_error} 
\end{equation}
and  the root of mean squared error (RMSE), defined as  $\mbox{RMSE}=\sqrt{\sum^k_{j=1}(\rho_j(\mathbf x^*) -\hat \rho_j(\mathbf x^*) )^2/k}$, follows 
\begin{equation}
     \mathbb P \big(\mbox{RMSE}>\delta\big) \leq \alpha. 
     \label{equ:P_RMSE}
  \end{equation}   
The detailed derivation of Eq. \eqref{equ:P_error} and \eqref{equ:P_RMSE} is given in Appendix \ref{sec:prob_ub}. Eqs. \eqref{equ:P_error} and \eqref{equ:P_RMSE} imply that the overall predictive error can be controlled in a probabilistic way under the maximum variance selection criterion in Eq. \eqref{equ:criterion_max}. 

In practice, the cut-off in (\ref{equ:criterion_max}) 
is too conservative, and we rarely see more than $\alpha$ samples larger than the claimed threshold. A less restrictive criterion to control the predictive error is by predicting any input set $\mathbf x^*$ such that 
\begin{equation}
    \delta^* := \sqrt{\frac{\sum_{j=1}^k\hat\sigma_j^2K^{**}}{k}} \leq \frac{\delta}{t_{\alpha/2}(n-q)}.
    \label{equ:criterion}
\end{equation}
We call the criterion in \eqref{equ:criterion} the average variance selection criterion. 
Eq. \eqref{equ:criterion} controls the predictive error of vectorized outputs (particle densities) in the average sense, while it does not guarantee the error control as in Eq. (\ref{equ:criterion_max}). In practice, one may select a large number of points for numerical solvers to run using criterion (\ref{equ:criterion_max}), thereby requiring a larger computational cost, while the predictive error control can be achieved using criterion (\ref{equ:criterion}) with much less training samples. 

To quantify whether the error is indeed controlled based on our threshold, we plot the empirical cumulative distribution function (CDF) of the absolute errors in Fig. \ref{fig:error} using the criterion (\ref{equ:criterion}) with different error bounds and probability tolerance thresholds $\alpha$ for all 4 classes of input functions considered herein.  
In the upper middle panel, for instance, the vertical solid line marks the chosen threshold $\delta=0.01$ and the purple dashed line labels the percentile corresponding to $1-\alpha=0.95$. On the x-axis, we mark the values of the 95th percentile of the absolute errors for each group of input, all of which are less than 0.01. 
In the remaining panels of Fig. \ref{fig:error}, the empirical CDFs of the absolute errors for various $\alpha$ and $\delta$ values are illustrated. In general, the smaller error bound or probability tolerance threshold, the smaller the predictive error is, while more samples are required for achieving this accuracy level. Thus, a balance between the computational cost and accuracy level is generally required. 


\subsection{Sequential update of the ALEC emulator} \label{subsec:update_derivation}
When calculating the predictive mean and predictive correlation in Eqs. (\ref{equ:gppredmean})-(\ref{equ:gppredcorrelation}), one needs to solve $\mathbf R^{-1}\mathbf z$, for an $n$-dimensional real-valued vector $\mathbf z\in \mathbb R^n$. In the PP-GP emulator \cite{Gu2016PPGaSP}, 
we implement the Cholesky decomposition of the correlation matrix $\mathbf R = \mathbf L\mathbf L^T$, and use the forward and backward substitution algorithms to compute $\mathbf R^{-1}\mathbf z$. 
In the ALEC emulator, whenever a sample $\mathbf x^* \in \mathbb R^p$ is added to the previous training set with $n$ samples, the PP-GP model needs to be updated accordingly. 
Directly applying this approach will require $\mathcal O(n^3)$ operations for computing the Cholesky decomposition each time when the sample size is $n$.  To reduce the computational cost, we introduce a new approach that takes only $\mathcal O(n^2)$ operations in performing the Cholesky decomposition. 

To update $\mathbf L$ with the addition of $\mathbf x^*$, we denote the correlation matrix of previous $n$ design elements as $\mathbf R_n=\mathbf L_n\mathbf L_n^T$, with $\mathbf L_n$ being the Cholesky decomposition of $\mathbf R_n$. Next, we write the updated correlation matrix $\mathbf R_{n+1}$:
\begin{equation*}
    \mathbf R_{n+1} = \begin{bmatrix}
    \mathbf R_n & \mathbf r_n(\mathbf x^*)\\
    \mathbf r_n^T(\mathbf x^*) & K(\mathbf x^*,\mathbf x^*)
    \end{bmatrix},
\end{equation*}
where $\mathbf r_n(\mathbf x^*) = [K(\mathbf x^*,\mathbf x_1),..., K(\mathbf x^*,\mathbf x_n)]^T$. In evaluating $\mathbf R_{n+1}=\mathbf L_{n+1}\mathbf L_{n+1}^T$, it is clear that the first $n$ rows of the lower triangular part of $\mathbf L_{n+1}$ is identical to the lower triangular part of $\mathbf L_n$ 
\begin{equation*}
    \mathbf L_{n+1} = \begin{bmatrix}
    \mathbf L_n & \mathbf 0_n\\
    \mathbf v_n & l_n
    \end{bmatrix}.
    \label{equ:updateL}
\end{equation*}
Therefore, only the last row of $\mathbf L_{n+1}$  needs to be computed:
\begin{align}
v_{n}(j) &= \frac{1}{L_n(j,j)}\Big( r_n(j)-\sum_{m=1}^{j-1}v_{n}(m)L_n(j,m)\Big), \label{equ:updateL1}\\
l_n &= \sqrt{K(\mathbf x^*,\mathbf x^*)-\sum_{m=1}^n v_{n}(m)^2},\label{equ:updateL2}
\end{align}
with $v_n(j)$ and $r_n(j)$ denoting  the $j$th element of $\mathbf v_n$ and $\mathbf r_n(\mathbf x^*)$, respectively, and $L_n(i,j)$ denoting the $(i,j)$ entry of $\mathbf L_n$. Eqs. \eqref{equ:updateL1} and \eqref{equ:updateL2} indicate that updating $\mathbf L_{n+1}$ requires $\mathcal O(n^2)$ computational operations, for  $i,j = 1,..., n$.
After obtaining the Cholesky decomposition $\mathbf L_{n+1}$ we can update other terms required in computing the predictive mean and variance.

\subsection{Algorithm and the computational cost}
\label{subsec:algorithm_cost}
We summarize the ALEC emulator in Algorithm \ref{alg:AL_GP}. Given the initial PP-GP emulator, predictive error bound $\delta$, and probability tolerance threshold $\alpha$, the ALEC emulator determines whether the particle densities can be reliably predicted for a new test input. The output contains the predictive particle densities for all test inputs and they are plugged into the energy functionals to compute the energy. 
Note that the range parameter in the kernel needs to be numerically estimated, which has the largest computational cost in the algorithm. When more training samples are added, the previously estimated range parameter $\hat\gamma$ may no longer be accurate if our initial model has a small training sample size. To compromise the computational cost and accuracy in predictions, the range parameter in the ALEC emulator is re-estimated for every 50 training samples, if the training size is less than 350. When the training size is large, the estimated range parameter does not change much. In general, the frequency of re-estimating the kernel parameter depends on the computational budget.  

Denote the initial training size as $n_{ini}$ and the number of augmented samples as $n_{aug}$. The computational cost of constructing the Cholesky decomposition of the initial ALEC emulator is $\mathcal O(n_{ini}^3)$, and the total computational order for updating the Cholesky decomposition is then $\mathcal O(\sum_{n=n_{ini}+1}^{n_{ini}+n_{aug}}n^2)$. For computing the predictive mean and predictive variance, it takes $\mathcal O(\sum_{n=n_{ini}+1}^{n_{ini}+n_{aug}}nk)$ and $\mathcal O(\sum_{n=n_{ini}+1}^{n_{ini}+n_{aug}}n^2k)$, respectively. 

The computational complexity  of the online updating approach reduces the cost compared to directing computing the Cholesky decomposition at each iteration, which costs $\mathcal O(\sum_{n=n_{ini}+1}^{n_{ini}+n_{aug}}n^3)$. The computational complexity is significantly faster than the time required to calculate the density using a numerical solver for computing the particle density at each test input. 
Indeed,  the largest computational cost of our approach comes from the numerical solver for the training input set, as will be seen in Fig. \ref{fig:time_bar}. Since only a small fraction of samples are used as the initial training and augmented sample, the computational cost of the ALEC emulator is much smaller than  running numerical solvers for each density. 



\begin{algorithm}[H]
\caption{ALEC emulator} \label{alg:AL_GP}
\begin{algorithmic}[1]
\Input{The number of test samples $n_t$, testing inputs $\mathcal X$, Cholesky factor $\mathbf L$ of $\mathbf R$ in the initial PP-GP model, estimated variance parameters $\hat{\bm\sigma}^2=[\hat{\sigma}_1^2,...,\hat{\sigma}_k^2]$ and mean parameters $\hat{\bm\theta}^2=[\hat{\theta}_1^2,...,\hat{\theta}_k^2]$ in the initial PP-GP model,  probability level $\alpha$ and  threshold  $\delta$ chosen by the user.
}
\State $n_{aug} \gets 0$;
\For{$i = 1$ to $n_t$}
\State Set $i$th test sample in $\mathcal X$ as $\mathbf x^*$ and calculate $\hat{\sigma}_j^2K^{**}$ for $j=1,...,k$ and $\delta^*$ via (\ref{equ:gppredcorrelation}) and (\ref{equ:criterion}), respectively;
\If {$\delta^* < \delta/t_{\alpha/2}(n-q)$}
\State Predict for the predictive distribution in (\ref{equ:predictiongp}) for this $\mathbf x^*$, and record predictive mean $\hat{\bm\rho}(\mathbf x^*)$ and scale $\hat{\bm\sigma}^2K^{**}$;
\Else
\State $n_{aug} \gets n_{aug}+1$;
\State Using the numerical solver to compute the corresponding output $\bm\rho(\mathbf x^*)$;
\If {the training size is a multiple of 50 \textbf{and} the training size $\leq 350$}
\State Rebuild the PP-GP emulator with retrained the range parameter $\gamma$ in Mat{\'e}rn kernel;
\Else
\State Update PP-GP emulator with the added input $\mathbf x^*$;
\State Update parameters $\hat{\bm\sigma}^2$ and $\hat{\bm\theta}^2$;
\EndIf
\EndIf
\EndFor
\Output{The number of augmented size $n_{aug}$, predicted mean, and variance of particle densities for remaining $n_t-n_{aug}$ samples.}
\end{algorithmic}
\end{algorithm}

\section{Numerical Results}
\label{sec:num_results}
Here we  evaluate the predictive performance of our surrogate model based on particle densities generated from four classes of external potentials (closed-forms shown in Appendix \ref{sec:close_form_external}) with the length of hard rods $a=1$ and the domain size $L=9$. The values of chemical external input and parameters in each external potential are changed when generating the data. We assume the external potential functions and chemical potentials  are used as inputs, while the parametric forms of the external potential functions are not known. We will test the predictive performance separately for each class of external potential functions, and jointly by combining all classes of  external potential functions, in Section \ref{subsec:separately} and Section \ref{subsec:jointly}, respectively. Section \ref{subsec:predicting_new} gives the predictive performance of our approach for predicting a new class of densities. 
The code and data used in this paper are publicly available: \url{https://github.com/UncertaintyQuantification/ALEC}. 

We record the out-of-sample root mean squared error (RMSE) of particle density and the grand potential of the test samples from
\begin{align*}
    {RMSE_\rho} &= \sqrt{\frac{\sum_{i=1}^{n^*}\sum_{j=1}^k(\rho_j(\mathbf x^*_i)-\hat \rho_j(\mathbf x^*_i))^2}{n^*k}}, \\
    RMSE_\Omega &= \sqrt{\frac{\sum_{i=1}^{n^*}(\Omega(\mathbf x^*_i)-\hat \Omega(\mathbf x^*_i))^2}{n^*}},
    \end{align*}
where $\hat \rho_j(\mathbf x^*_i)$ is the predicted density of $i$th hold out sample at $j$th coordinate, and $\Omega(\mathbf x^*_i)$ and $\hat \Omega(\mathbf x^*_i)$ are the predicted and true grand potential for $i$th hold out sample, respectively.
Other criteria, such as the proportion of the test samples covered in the 95\% predictive intervals and the average length of the predictive intervals, are provided in supplementary materials. Note that we only compute the predictive error on those inputs we predict, not including those initial samples or the augmented samples computed by the numerical solver. 
    
We generate 2000 samples for each group with LHS, and compare our surrogate model with three other commonly used approaches. The first two approaches use conventional random sample (RS) designs, 
where we randomly draw $n_{train}$ samples from the LHS samples as the training dataset and use the remaining samples as testing data.
We have two distinct training sample sizes for RS samples. The sample size in the first RS sample (RS1) is specified as the initial training size in ALEC, while the sample size in the second RS sample (RS2) is specified as the final training size in ALEC. For predicting any test input, the training sample size in active learning is neither smaller than the training sample size in RS1 nor larger than the training sample size in RS1. We have also implemented the active learning algorithm with D-optimality  \cite{podryabinkin2017active}, where the details are given in Appendix \ref{sec:d-optimality}.  



\subsection{Particle density and grand potential prediction for each group of external potential field separately}
\label{subsec:separately}
\begin{figure}
    \centering
    \includegraphics[width=9cm]{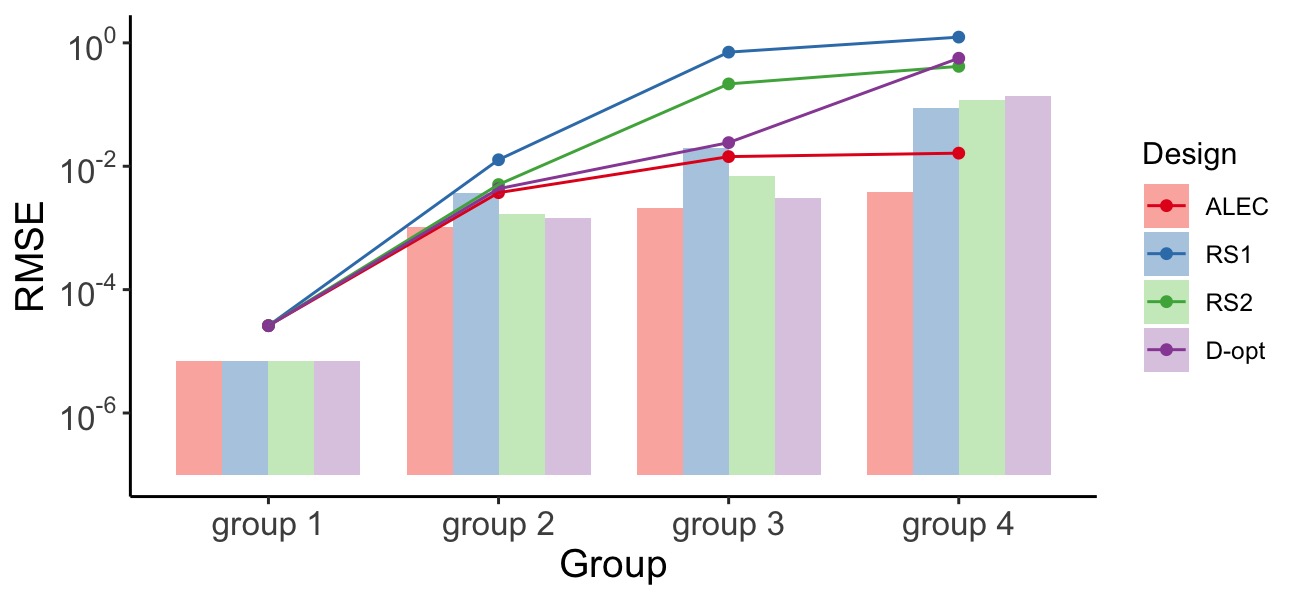}
    \caption{The bar plots and line plots 
    show the out-of-sample RMSE of density $\rho$,  and RMSE of grand potential $\Omega$, respectively. The ALEC emulator is compared with other methods with three other designs: RS1, RS2, and D-opt (Active learning with D-optimality). For the ALEC emulator, we use $\delta=0.01$ and $\alpha=0.05$ as the threshold, and the augmented sizes for groups 1-4 are 0, 7, 21, and 504, respectively. In D-opt, the augmented sizes for groups 1-4 are 0, 7, 21, and 582, respectively. The number of training sample sizes from RS1 and RS2 are the same as the initial sample size and terminal sample size in ALEC. }
    \label{fig:RMSE_model_separately}
\end{figure}

\begin{figure*}[t]
    \centering
    \includegraphics[width=16.5cm]{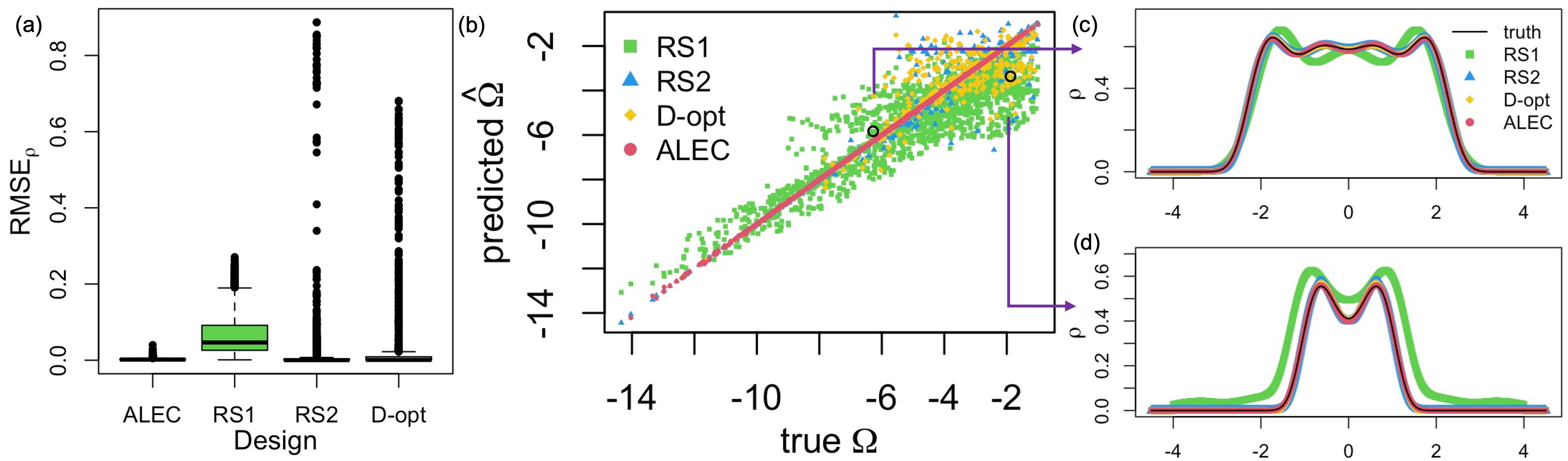}
    \caption{A comparison of four surrogate models for predicting the group 4 functions: Active learning with error control (ALEC), random sampling with the number of training points as the initial number of training points in ALEC (RS1) and with the number of training points as the final number of training points in ALEC (RS2), and active learning with D-optimality (D-opt). (a) Boxplots of predictive RMSE of each predicted density in the testing dataset. (b) Predicted grand  potential $\hat\Omega$ vs. the truth $\Omega$. (c)-(d) True and predicted densities for two selected density profiles.}
    \label{fig:group4_pred} 
\end{figure*}

We first test different approaches for four families of functional inputs separately. The density profiles are initially trained using PP-GP with $n_{ini}=20$ inputs per group. Then following our ALEC algorithm, the surrogate model is sequentially updated by augmenting the train data with samples selected by the average variance selection criterion in Eq. (\ref{equ:criterion}). After obtaining the predicted density $\hat{\rho}$ for the remaining test samples, we predict the grand potential $\Omega$ by plugging $\hat{\rho}$ into the energy functional in Eq.  \eqref{equ:close_Omega} in Appendix \ref{sec:close_form_cdft}.

Figure \ref{fig:RMSE_model_separately} displays the overall performance of different surrogate models (more detailed results can be found in the table in the supplementary materials). Here the threshold in the ALEC algorithm is set as $\delta=0.01$ and $\alpha=0.05$, representing a strategy for having less than $5\%$ of the coordinate to have an absolute error of density prediction larger than $0.01$. The first group of external potentials has the least flexibility, as the only change in the input is $\mu$. Thus, predicting the density profile for the first group of functions is the easiest. Models with only 20 inputs in the conventional RS design can predict particle densities relatively well, and no augmented sample is needed in the ALEC emulator. For the other three functional groups, the active learning algorithm has the smallest predictive RMSE on density $\rho$ and grand potential $\Omega$ among all four designs. 
It is worth noting that the number of training inputs from active learning is always not larger than that in the 
RS2 design, but the predictive error by active learning is much smaller than the one by RS2. This is because the active learning strategy accurately identifies test samples hard to be predicted, and a numerical solver is called for computing these samples. Thus the computing budget is more efficiently spent with respect to the accuracy of predictions.  

Figure \ref{fig:group4_pred}  presents more detailed prediction results of the particle densities and grand potentials for the fourth class of external potentials. The boxplot in Fig. \ref{fig:group4_pred}(a) gives the out-of-sample RMSE for predicting the particle densities. Many densities are inaccurately predicted based on the samples from RS1, RS2, or D-opt, resulting in large errors in predicting the grand potential  (Fig. \ref{fig:group4_pred}(b)). In comparison, the ALEC emulator identifies the samples that are difficult to be predicted and it uses them to enhance the emulator, thus providing accurate predictions for the remaining test samples. Although the total computational cost of the ALEC is similar to the methods with  RS2  and D-opt schemes, the ALEC emulator results in better predictive accuracy. 

\subsection{Predicting particle densities and grand potentials for all groups of external potentials}
\label{subsec:jointly}
In this subsection, we describe a more challenging scenario, where we aim to predict four groups of samples together. We assume that only the chemical potential and the external potential are used as input, 
while the parametric forms of the external potential are not known. The  density profile and grand potential are  harder to be predicted accurately in this case, as the input contains more degrees of freedom,  leading to larger variability of the density and energy functionals,  compared to the ones from any single class of input function alone.

Since the inputs are a random sample from four classes of external potentials, we use the first $n_{ini} = 80$ samples  as initial inputs to build the ALEC emulator. When a test sample comes, we  use the criterion from Eq.  (\ref{equ:criterion})  to determine whether the test sample can be accurately predicted by the ALEC  emulator   or a numerical solver will be needed for solving the system directly. 

\begin{figure}
    \centering
    \includegraphics[width=9cm]{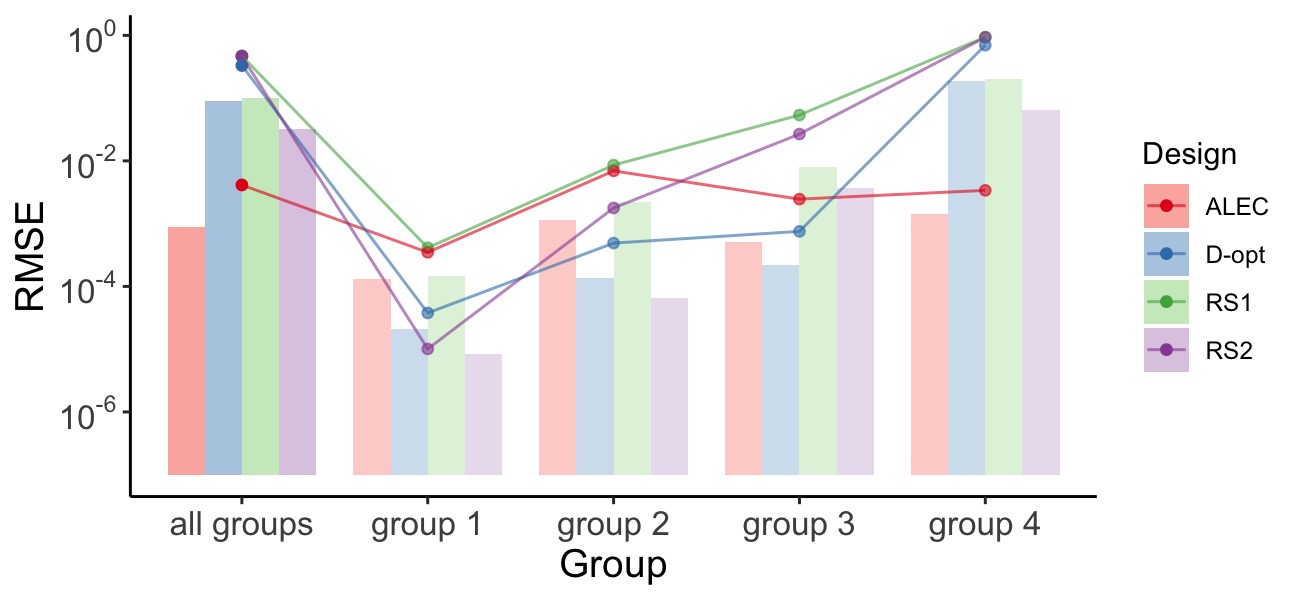}
    \caption{Out-of-sample RMSE of particle density $\rho$ (bar plots) and grand potential $\Omega$ (line plots) for active learning with error control, RS1 design, RS2 design and active learning with D-optimality. The overall RMSE for combined all groups  (solid bars and dots) and the RMSE of each individual group (shadow bars and dots) are both recorded.
    ALEC contains 80 inputs as initial training size, and the total number of augmented samples in the ALEC emulator ($\delta = 0.01$, $\alpha=0.05$) is 873, with 0, 1, 58, and 814 augmented samples in groups 1 to 4, respectively. 
    In D-opt, the total number of augmented samples is 935, and the augmented sizes for groups 1-4 are 59, 73, 393, and 410, respectively. RS1 contains a total of the first 80 training samples after uniformly mixing all four groups of density together.  RS2 contains the first 953 training inputs,  with 245, 221, 238, and 249 for groups 1-4, respectively. Since the ALEC emulator identifies the hardest region (group 4) to be predicted, thereby augmenting most  samples from group 4,   it has the smallest predictive error overall and predictive errors of all groups are controlled.  
    }
    \label{fig:RMSE_model_together}
\end{figure}

We show the emulator's overall performance in the leftmost group (named as all groups) in Fig. \ref{fig:RMSE_model_together} (detailed results are displayed in the supplementary materials). On average, the ALEC emulator  performs much better, where the predictive RMSE of both particle densities and grand potentials is at least one magnitude smaller than all other approaches. A total of 873 training samples are added through the active learning algorithm, which is more than the total number of training samples added for building the emulator separately. This is because having a larger input space containing all 4 classes of external potentials increases the difficulty of learning densities, requiring more samples given the same threshold of the error. 
Note that the good predictive performance by ALEC is achieved using a similar or smaller sample size in training the emulator compared to the D-opt and RS2 approaches, by efficiently allocating the computing budget on computing particle densities hard to be predicted, and accurately predicting the rest of the densities  with a negligible computational cost.


When examining the performance of different approaches  in each function group, we found that the RMSE of particle densities from ALEC is homogenous across groups, while RS2 and D-opt approaches have smaller predictive errors for the first two groups, but have much worse performance for the fourth group. This is because RS2 and D-opt select more points from the first two groups to train the emulator, thereby  yielding a similar or better result than ALEC in these two groups. However, the error from the first two groups can be controlled below the threshold of $0.01$ for more than 95\% of the test samples  with  no augmented samples or only a small number of augmented samples. In the ALEC emulator, for instance, the augmented samples from groups 1 and 2 are 0 and 1, respectively. On the contrary, the particle densities and grand potential energy from group 4 are the hardest to be predicted. This is automatically identified by ALEC, and the most computing budget (814 augmented samples) is spent on the fourth group to ensure the predictive errors of most samples are controlled below the threshold.  
This is exactly what we intend to accomplish. The ALEC can identify input regions that are difficult to predict, and then put more computing resources on those regions to control the predictive error.  Consequently, the predictive error is controlled below the threshold for all subclasses, even if we do not know the labels of subclasses from the test samples, as all input classes are combined for predictions. 

Another interesting finding is that the additional training samples from other input functional classes do not substantially improve the predictive performance for a specific input class. This is because the inputs from one class are distant from another class, as shown in Fig. \ref{fig:functional_learn}(b), and thus the correlation between different input classes is small. 
It may be of interest to automatically identify the cluster in the training and split the training sample when constructing the emulator. This is useful as splitting the samples can reduce the computational complexity of the  emulator.
To explore this idea, 
we utilized a simple input decomposition technique. When the size of a cluster reaches 400, we divide it into two sub-clusters using the K-nearest neighbors (KNN) algorithm \cite{fix1989discriminatory}. In this approach, the number of augmented samples and predictive RMSE of particle densities are 779 and $9.73\times 10^{-4}$, respectively. It achieves similar accuracy with 94 fewer training samples than the ALEC  algorithm without input decomposition.
 Developing a more comprehensive way for input decomposition approach can further enhance the efficiency and accuracy of  the surrogate model.



\begin{figure}[t]
    \centering
    \includegraphics[width=9cm]{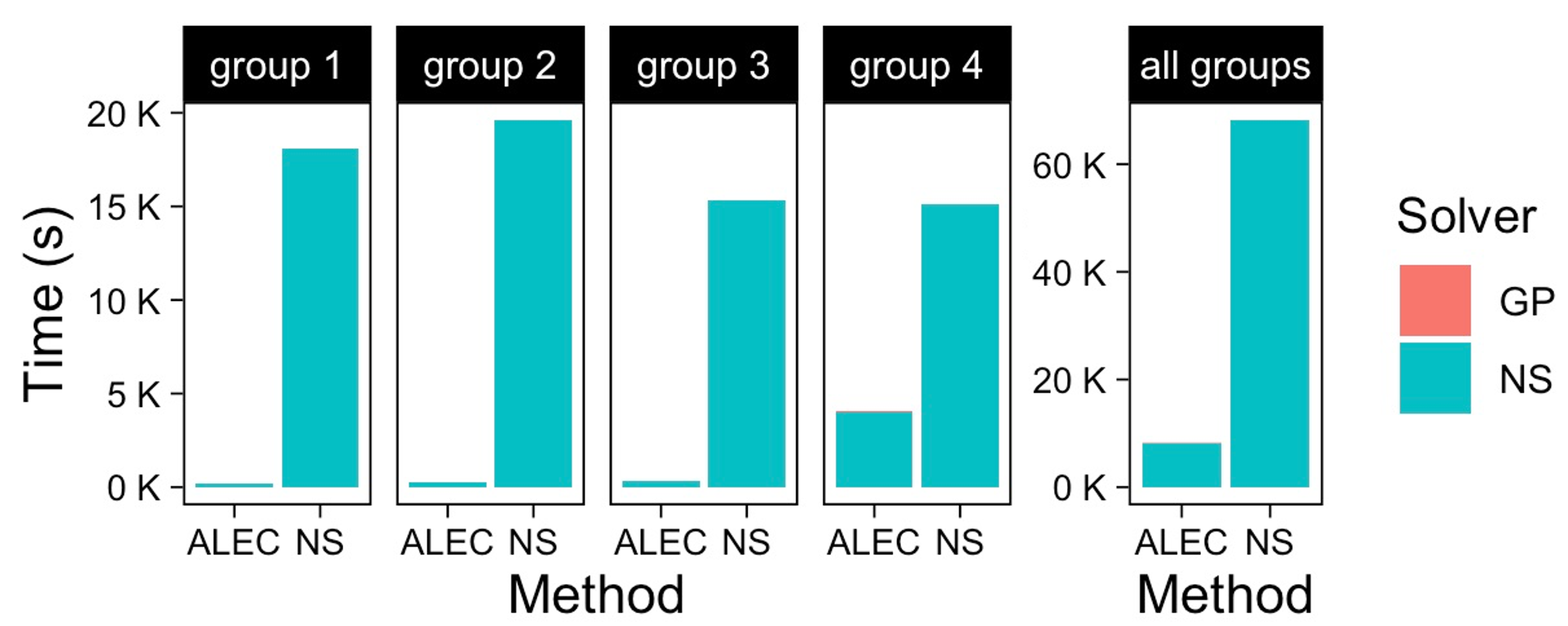}
    \caption{The running time for each function group using the ALEC algorithm and using numerical solver (NS). As the running time for building the GP model in ALEC is very small, the bars for GP are nearly invisible. Other approaches, such as  RS2 and D-opt, have similar computational costs as ALEC as the training and augmented sample sizes are similar.}
    \label{fig:time_bar}
\end{figure}

Fig. \ref{fig:time_bar} compares the computational time required in the ALEC approach and in calling a numerical solver (NS) each time. 
The largest computational cost in the ALEC algorithm comes from calling the NS for computing the particle densities of the initial and augmented training samples, while the time in constructing the ALEC emulator and making predictions is negligible, as the training sample size is not large. 
Since the ALEC emulator  only requires a fraction of the samples to be numerically solved, it is much faster than running a numerical solver for each test sample. Note that RS2 and D-opt compared herein 
have a similar computational cost as the ALEC emulator, as the number of density profiles to be solved by NS is similar, yet the predictive error was not controlled in those approaches. 
\label{sec:new_Vext}
\begin{figure*}[t]
    \centering
    \includegraphics[width=16.5cm]{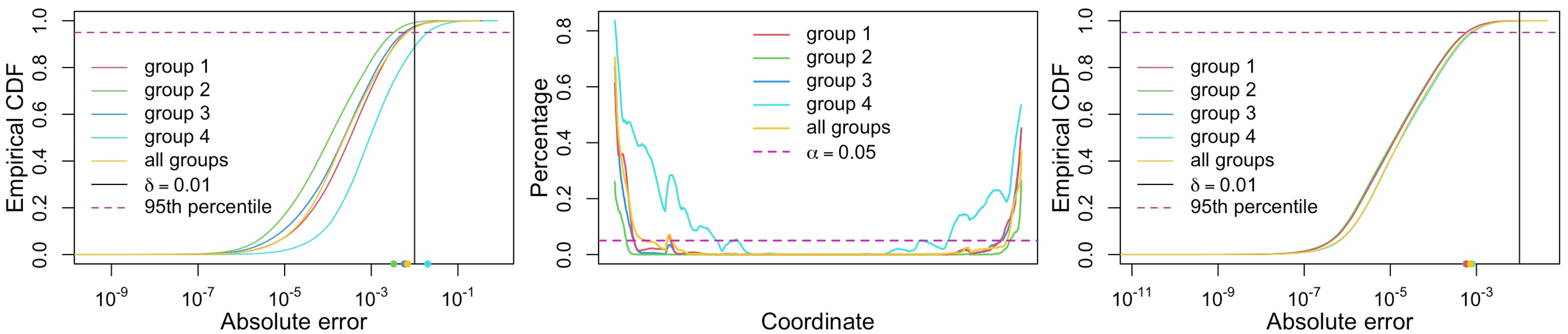}
    \caption{Prediction results from the ALEC emulator on a new class of external potentials where  ALEC  models are trained by the previous four functional groups separately and by all functional groups together. $\delta = 0.01$ and $\alpha=0.05$ are used. Left panel: the empirical CDF of absolute error using the average variance selection criterion  in Eq. \eqref{equ:criterion}. The number of augmented samples using the model for groups 1-4 and all groups combined are 92, 122, 67, 48, and 15, respectively.  Middle panel: the percentage that the absolute difference is greater than $\delta$ for each coordinate, i.e. $\sum_{i=1}^{n*}\mathbbm{1}\{\rho_j(\mathbf x_i^*)-\hat\rho_j(\mathbf x_i^*) > \delta\}/n^*, j=1, ..., k$. Right panel: the empirical CDF of absolute error using  the maximum variance selection criterion in Eq. \eqref{equ:criterion_max}. The number of augmented samples using the model for groups 1-4 and all groups combined are 408, 395, 384, 362, and 263, respectively.}
    \label{fig:pred_mixed23}
\end{figure*}
\subsection{Predicting a new class of external potential}
\label{subsec:predicting_new}

In this section, we examine the performance of the ALEC approach to predict inputs from a new functional form that is not in the training data set. 
To test this case, in addition to the existing 8000 samples, we generated 2000 new samples based on the weighted inputs between groups 2 and 3 as explained in Appendix \ref{sec:close_form_external}. We test the  performance of the ALEC emulator for predicting the new input class, 
starting with the final GP emulator developed for groups 1-4 and all groups combined. 

The left panel of Fig. \ref{fig:pred_mixed23} displays empirical CDF plots of the absolute errors of the ALEC emulator using the average variance selection criterion in Eq. \eqref{equ:criterion}. First, the predictive error of most of the test samples in the new group is small, for any initial emulator built upon other groups. Less than $\alpha=5\%$ of the absolute errors are greater than the chosen error bound $\delta=0.01$ for models using groups 1-3 and all groups combined. When we start from the initial  emulator built from group 4 input, $11.2\%$ of predictive errors are greater than $\delta=0.01$, which is larger than the probability tolerance threshold   $\alpha=5\%$. A close examination reveals that the percentage of the absolute error greater than $\delta$ is typically large for coordinates at the tail of the particle densities (Fig. \ref{fig:pred_mixed23} middle panel), because of a  rapid change in density at both ends, making it distinct from existing densities in group 4. The difference in the particle densities in the new class and group 4 increases the model discrepancy and makes the uncertainty in predictions harder  to be quantified. 
Among the 11\% of the test samples with RMSE larger than the threshold, however, most RMSEs are still very close to the threshold value, making this criterion ideal if the computational budget is not large. 

On the other hand, 
the right panel of Fig. \ref{fig:pred_mixed23}  shows the CDF of the RMSEs based on the maximum variance selection criterion  in Eq. (\ref{equ:criterion_max}). Much less than $\alpha$ probability   of the predictive error exceeds the threshold $\delta$ for all the cases shown here, as the  maximum variance selection criterion is conservative, yielding more augmented samples and higher  predictive accuracy. When the model discrepancy is large, one may use the maximum variance selection criterion to achieve a higher level of accuracy in predictions.


\section{Conclusion}
\label{sec:conclusion}
In this work, we propose the ALEC emulator, an active learning framework for controlling the predictive error of approximating computationally expensive functionals, based on the internal uncertainty assessment of the statistical surrogate model. 
This result has a probabilistic interpretation and it is fully automatic. 
We test the ALEC emulator on cDFT calculations of
one-dimensional hard rods with different chemical potentials and external potentials.  Through the numerical simulations of the density profiles and grand potentials for different groups of functions, we show that ALEC outperforms the conventional random sample designs, as well as the active learning approach with the D-optimality criterion in terms of prediction accuracy. 
With a fraction of the computational cost of running a numerical solver for each test input, we are able to predict functionals accurately with a controlled predictive error using the ALEC emulator. The ALEC emulator is designed as a reliable building block to be integrated for solving challenging simulation tasks, such as predicting expensive cDFT calculations on functional input, since the predictive error of the ALEC emulator is controlled for the test samples. 
 


There are several future research directions. First,  the ALEC emulator was built upon the PP-GP surrogate model for predicting vectorized output, while the approach can be  generalized to using other surrogate models with uncertainty assessment in predictions. Physical symmetries, for instance, can be incorporated to define a new descriptor and distance metric to constrain the surrogate model. For instance, a common way to ensure translational invariance is to use pairwise distances of input, instead of the input function itself. Existing functionals found in physics \cite{ma2022evolving}  may be integrated as the mean of the emulator to improve its accuracy. Furthermore, dimension reduction of the input space through orthogonal representation can be helpful for reducing computational costs and improving the accuracy of predicting systems on 3D coordinates. Potential approaches include basis representation for reducing the dimensionality of the output space \cite{brockherde2017bypassing} and orthogonal projection of the inputs by maintaining large gradients of outcomes for reducing the dimensionality of the input space   \cite{constantine2014active}.  
Finally, one may jointly model the grand potential energy and densities for predicting both quantities. This is particularly useful for applications with complex molecular systems, where computing the potential energies from particle densities is  expensive. 
 

\section*{Supplementary materials}
The supplementary materials contain the derivation of predictive t-distribution in  Eq. (\ref{equ:predictiongp}), and additional prediction results of particle density and grand potential. 

\section*{Acknowledgements}

This study is partially supported by the National Science (NSF) Foundation under Award No. DMS-2053423. X.F. acknowledged  partial support from the NSF BioPACIFIC Materials Innovation Platform (MIP) under Award No. DMR-1933487 and the UCSB academic senate faculty research grants program. J.W. acknowledged partial support by the NSF Harnessing the Data Revolution (HDR) Big Ideas Program under Grant No. NSF 1940118.  

X.F. and M.G. contributed equally to this work.

\section*{Data Availability Statement}
The data that support the findings of this study are available from the corresponding author upon reasonable request.

\appendix
\section{Analytical forms in cDFT} \label{sec:close_form_cdft}
The closed-form expressions of $\delta F_{ex}/\delta\rho$, $\Omega$, and $ F_{ex}$ for 1D HR \cite{percus1976equilibrium,tarazona2008density} are summarized here:
\begin{align}
    \frac{\delta \beta F_{ex}[\rho]}{\delta\rho} =& \int_x^{x+a}\frac{\rho(z)}{1-\int_{z-a}^z\rho(y)dy}dz-\nonumber\\
    &\qquad\quad\ln\Big(1-\int_{x-a}^x\rho(y)dy\Big)
    \label{equ:deriv_F_ex}\\
    \beta\Omega[\rho] =&-\int_{-\infty}^\infty \frac{\rho(x-a)}{1-\int_{x-a}^{x}\rho(y)dy}dx,
    \label{equ:close_Omega} 
    \\
     \beta F_{ex}[\rho] =& - \int \rho(x)\ln\Big(1-\int_{x-a}^x\rho(y)dy\Big)dx, \label{equ:close_F_ex}
\end{align}
where $a$ is the length of each hard rod.

\section{External potentials and chemical potentials}
\label{sec:close_form_external}
\begin{figure*}[ht]
    \centering
    \includegraphics[width=17cm]{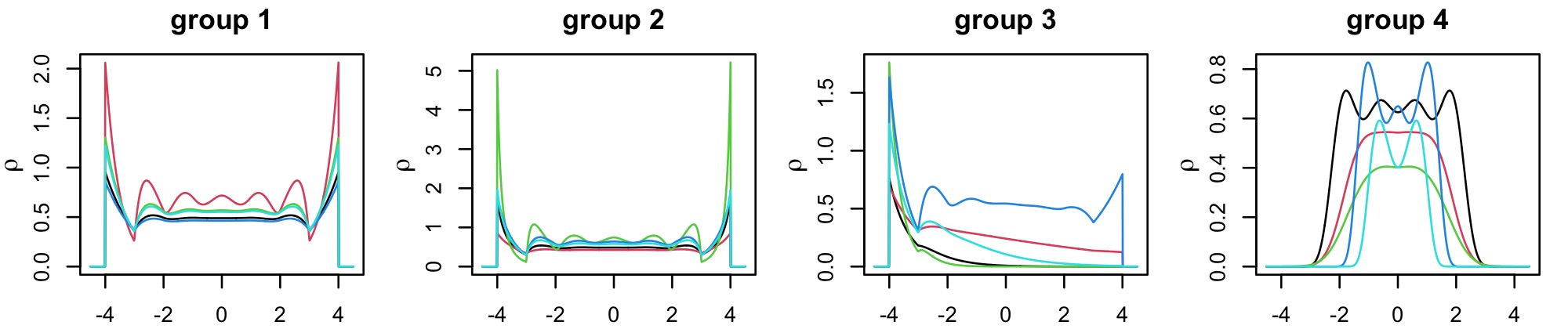}
    \caption{Each plot gives 5 examples of generated densities from external potential in \ref{equ:zero_inf}-\ref{equ:power}.}
    \label{fig:density_example}
\end{figure*}
Let $L$ be the distance between two hard walls, and $a$ is the length of the hard rods (HR). We consider the following four parametric forms of the external potential in this paper.

The first functional form applies to uniform hard rods confined between hard walls \cite{vanderlick1989statistical}:
\begin{equation}\label{equ:zero_inf}
    V^{ext}(s) = \left\{
    \begin{array}{ll}
        \infty, & s<a/2, s>L-a/2, \\
        0, & a/2<s<L-a/2.
    \end{array}
    \right.
\end{equation}
In the second functional form, the hard rods experience  a van der Waals-like attraction from the hard walls
\begin{equation}\label{equ:binary}
    V^{ext}(s) = \left\{
    \begin{array}{ll}
        \infty, & s<a/2, s>L-a/2, \\
        \\
        -\epsilon\{(\frac{a}{s+a/2})^3+&(\frac{a}{L+a/2-s})^3\}, \\
        & a/2<s<L-a/2.
    \end{array}
    \right.
\end{equation}
Eq.\eqref{equ:binary} is adopted from \cite{vanderlick1989statistical}. With fixed $a$ and $L$, $\epsilon$ is a parameter in the range of $\epsilon\in(0.1,2.2)$. Note that when $\epsilon = 0$, we have the same potential as the first one.

The third functional form consists of hard-wall potentials and a repulsion linearly proportional to the distance 
\begin{equation}\label{equ:grav}
   V^{ext}(s) = \left\{
    \begin{array}{ll}
        \infty, & s<a/2, s>L-a/2, \\
        m_gs, & a/2<s<L-a/2.
    \end{array}
    \right.
\end{equation}
Intuitively, the linear form mimics the gravitational potential.\cite{bakhti2015monodisperse} This is the only asymmetric potential considered in this work. In training the machine-learning models, the range of $m_g$ is chosen to be $(0.1, 3)$.  

The power-law external potential \cite{bakhti2015monodisperse} is the fourth functional class considereed in this work. This external potential has more flexibility as it has three parameters, $u_0, x_0$, and $a_0$ that can be used to modify the range of interactions:
\begin{equation}\label{equ:power}
    V^{ext}(s) = \left\{
    \begin{array}{ll}
        \infty, & s<a/2, s>L-a/2, \\
        u_0\big|\frac{s-L/2}{x_0}\big|^{a_0}, & a/2<s<L-a/2,
    \end{array}
    \right.
\end{equation}
where $a_0>0$. In this work, we use $u_0\in(1,3), x_0\in(1,3)$ and $a_0\in(2,5)$. 
Fig. \ref{fig:density_example} shows five examples of generated densities for each external potential.

In section \ref{sec:new_Vext}, the performance of our strategy is evaluated on a new set of densities. The external potential of these densities is generated from the weighted average of (\ref{equ:binary}) and (\ref{equ:grav}):
\begin{equation}
   V^{ext}(s) = \left\{
    \begin{array}{ll}
        \infty, & s<a/2, s>L-a/2, \\
        \\
        -w\epsilon\{(\frac{a}{s+a/2})^3+&(\frac{a}{L+a/2-s})^3\}+(1-w)m_gs, \\
        & a/2<s<L-a/2,
    \end{array}
    \right.
\end{equation}
where $w \in (0,1)$.


\section{Derivation of probabilistic upper bounds}
\label{sec:prob_ub}
First, we derive the probabilistic upper bound for the absolute difference of densities. For a fixed coordinate $j$, based on the predictive distribution in (\ref{equ:predictiongp}), we have $\mathbb P\left(|\rho_j(\mathbf x^*)-\hat{\rho}_j(\mathbf x^*)|>\sqrt{\hat{\sigma}_j^2K^{**}}t_{\alpha/2}(n-q)\right) = \alpha$. Since $\rho_j(\mathbf x^*)$ is predicted for sample $\mathbf x^*$ only if $\sqrt{\hat\sigma^2_j K^{**}}\leq \delta_{j}/t_{\alpha/2}(n-q)$,
the probability of predictive error larger than error bound $\delta_j$ follows
\begin{align*}
    & \mathbb P\left(|\rho_j(\mathbf x^*)-\hat{\rho}_j(\mathbf x^*)|>\delta_j\right)\\
    =& \mathbb P\left(|\rho_j(\mathbf x^*)-\hat{\rho}_j(\mathbf x^*)|>\frac{\delta_{j}}{t_{\alpha/2}(n-q)}t_{\alpha/2}(n-q)\right)\\
    \leq& \mathbb P\left(|\rho_j(\mathbf x^*)-\hat{\rho}_j(\mathbf x^*)|>\sqrt{\hat{\sigma}_j^2K^{**}}t_{\alpha/2}(n-q)\right) =\alpha. 
\end{align*}

Next, let us consider controlling the grid-level predictive error using the maximum variance selection  criterion in (\ref{equ:criterion_max}), where $\bm\rho(\mathbf x^*)$ is predicted for $\mathbf x^*$  only if $\sqrt{\max_j\{\hat\sigma^2_j K^{**}\}}\leq \delta/t_{\alpha/2}(n-q)$. 
Then for any $j$, $j=1,...,k$, 
\begin{align*}
    & \mathbb P\left(|\rho_j(\mathbf x^*)-\hat{\rho}_j(\mathbf x^*)|>\delta\right)\\
    \leq& \mathbb P\left(|\rho_j(\mathbf x^*)-\hat{\rho}_j(\mathbf x^*)|>\sqrt{\max_j\hat{\sigma}_j^2K^{**}}t_{\alpha/2}(n-q)\right) \\
        \leq& \mathbb P\left(|\rho_j(\mathbf x^*)-\hat{\rho}_j(\mathbf x^*)|>\sqrt{\hat{\sigma}_j^2K^{**}}t_{\alpha/2}(n-q)\right)=\alpha, 
 \end{align*}
from which equation (\ref{equ:P_error}) is proved. 

Third, let us derive the probabilistic bound for root mean squared error (RMSE) under the maximum variance selection  criterion in Eq. (\ref{equ:criterion_max}), where $\mbox{RMSE}=\sqrt{\sum^k_{j=1}(\rho_j(\mathbf x^*) -\hat \rho_j(\mathbf x^*) )^2/k}$. For $j=1,...,k$, we have 
\begin{align*}
    &\Prob(\mbox{RMSE}>\delta) \\
    =& \Prob\left(\sum^k_{j=1}\frac{(\rho_j(\mathbf x^*) -\hat \rho_j(\mathbf x^*) )^2}{k}>\delta^2\right)\\
    \leq&\Prob\left(\max_j (\rho_j(\mathbf x^*) -\hat \rho_j(\mathbf x^*) )^2 >\delta^2\right)\\
        \leq&\Prob\left(\max_j (\rho_j(\mathbf x^*) -\hat \rho_j(\mathbf x^*) )^2 >\max_j\hat{\sigma}_j^2K^{**}t^2_{\alpha/2}(n-q)\right),
\end{align*}
where the third inequality follows from the condition in Eq.  (\ref{equ:criterion_max}). 
Supposing the left hand side inside the probability is maximized when $j=j^*$,  then we have 
\begin{align*}
    &\Prob(\mbox{RMSE}>\delta) \\
    \leq&\Prob\left((\rho_{j^*}(\mathbf x^*) -\hat \rho_{j^*}(\mathbf x^*) )^2  >\max_j\hat{\sigma}_j^2K^{**}t^2_{\alpha/2}(n-q)\right), \\
        \leq&\Prob\left((\rho_{j^*}(\mathbf x^*) -\hat \rho_{j^*}(\mathbf x^*) )^2 >\hat{\sigma}_{j^*}^2K^{**}t^2_{\alpha/2}(n-q)\right) =\alpha,
    \end{align*}
    from which equation \eqref{equ:P_RMSE} is proved.



Lastly, we give the upper bound for the expectation of mean squared error, $\mbox{MSE}=\sum^k_{j=1}(\rho_j(\mathbf x^*) -\hat \rho_j(\mathbf x^*) )^2/k$, under the average variance selection criterion (Eq. (\ref{equ:criterion})),
\begin{align*}
    \mathbb E_{\bm \rho(\mathbf x^*)} (\mbox{MSE}) &= \mathbb E_{\bm \rho(\mathbf x^*)}\left[\sum^k_{j=1}\frac{(\rho_j(\mathbf x^*) -\hat \rho_j(\mathbf x^*) )^2}{k}\right]\\
    &= \sum^k_{j=1}\frac{\hat\sigma_j^2K^{**}}{k}\frac{n-q}{n-q-2}\\
    &\leq\frac{\delta^2}{t_{\alpha/2}^2(n-q)}\frac{n-q}{n-q-2}.
\end{align*}
For moderately large $n$,  and small $\alpha$, we have $\mathbb E_{\bm \rho(\mathbf x^*)} (\mbox{MSE}) \leq \delta^2$. For instance, for any $n\geq 20$ and constant mean basis ($q=1$), $\mathbb E_{\bm \rho(\mathbf x^*)} (\mbox{MSE}) \leq \delta^2$ for any $\alpha\leq 0.2$.

\section{Details of the D-optimality criterion}
\label{sec:d-optimality}
Here we discuss the details of the D-optimality criterion. Consider the correlation function $K(\mathbf x,\mathbf x_i), i=1,..., n,$ with each training set sample serving as a set of basis functions. Then, by definition, these basis functions of the training dataset form the correlation matrix $\mathbf R$, and the basis functions with any testing sample $\mathbf x^*$ gives $\mathbf r^T(\mathbf x^*)$. Define the row vector $\mathbf c = \mathbf r^T(\mathbf x^*)\mathbf R^{-1}$ for each sample $\mathbf x^*$. 
Set the threshold $c_{th}\geq 1$ for the maximum absolute value in $\mathbf c$. If the uncertainty estimate $c_{max}:=\max|c_i|>c_{th}$, 
a numerical solver will be called to solve the system and the outcomes at $\mathbf x^*$ will be added to the training set. Otherwise,  the emulator will be used to predict the test input $\mathbf x^*$.  
Since the D-optimality criterion is not directly related to the predictive error, 
selecting threshold $c_{th}$ may be performed in a case-by-case manner. 
In order to make a comparison with our method, the threshold is adjusted to make the number of augmented samples selected from D-optimality at least as large as the one in our strategy. For more discussion about the D-optimality criterion in emulating molecular simulations, we refer the readers to  \cite{podryabinkin2017active} and section 15.2.3 in \cite{schutt2020machine}.

\renewcommand\refname{References}
\bibliography{References_2021}

\end{document}